\documentclass[runningheads]{llncs}
\usepackage{graphicx}
\usepackage[position=top]{subfig} 
\usepackage[table,xcdraw]{xcolor}
\usepackage{amsmath,amssymb}
\usepackage{hyperref}
\hypersetup{colorlinks=true}

\usepackage{cite}

\usepackage{array}
\newcolumntype{P}[1]{>{\centering\arraybackslash}p{#1}}
\newlength\savewidth\newcommand\shline{\noalign{\global\savewidth\arrayrulewidth
  \global\arrayrulewidth 0.8pt}\hline\noalign{\global\arrayrulewidth\savewidth}}

\newenvironment{jlblue}{\color{blue}}{}

\newenvironment{jlmagenta}{\color{magenta}}{}

\usepackage{soul}
\newcommand{\jlreplace}[2]{{\jlblue #2}}

\newcommand{\drmgreplace}[2]{{\jlblue #2}}
\newcommand{\drmgadd}[1]{{\jlmagenta #1}}

\newcommand{\eg}{\mbox{e.g.,\ }}

\newcommand{\ie}{\mbox{i.e.,\ }}

\begin{document}
\title{Seeking an Optimal Approach for Computer-Aided Pulmonary Embolism Detection}% \thanks{Supported by organization x.}}
\titlerunning{Seeking an Optimal Approach for Computer-Aided PE Detection}
% \author{Anonymous} 
% \and

\author{
    Nahid Ul Islam\inst{1} \and
    Shiv Gehlot\inst{1} \and
    Zongwei Zhou\inst{1} \and \\
    Michael B Gotway\inst{2} \and
    Jianming Liang\inst{1}
}

\authorrunning{N. Islam, et al.}

\institute{
    Arizona State University, Tempe, AZ 85281, USA \\
    \email{\{nuislam,sgehlot,zongweiz,jianming.liang\}@asu.edu} \and
    Mayo Clinic, Scottsdale, AZ 85259, USA 
    \email{Gotway.Michael@mayo.edu}
}

\maketitle              % typeset the header of the contribution
\begin{abstract}

Pulmonary embolism (PE) represents a thrombus (``blood clot''), usually originating from a lower extremity vein, that travels to the blood vessels in the lung, causing vascular obstruction and in some patients, death. This disorder is commonly diagnosed using CT pulmonary angiography (CTPA). Deep learning holds great promise for the computer-aided CTPA diagnosis (CAD) of PE. However, numerous competing methods for a given task in the deep learning literature exist, causing great confusion regarding the development of a CAD PE system. To address this confusion, we present a comprehensive analysis of competing deep learning methods applicable to PE diagnosis using CTPA at the both image and exam levels. At the image level, we compare convolutional neural networks (CNNs) with vision transformers, and  contrast self-supervised learning (SSL) with supervised learning, followed by an evaluation of transfer learning compared with training from scratch. At the exam level, we focus on  comparing conventional classification (CC) with multiple instance learning (MIL). Our extensive experiments consistently show: (1) transfer learning consistently boosts performance despite differences between natural images and CT scans, (2) transfer learning with SSL surpasses its supervised counterparts; (3) CNNs outperform vision transformers, which otherwise show satisfactory performance; and (4) CC is, surprisingly, superior to MIL. Compared with the state of the art, our optimal approach provides an AUC gain of 0.2\% and 1.05\% for image-level and exam-level, respectively. 

\keywords{Pulmonary Embolism \and CNNs \and Vision Transformers \and Transfer Learning  \and Self-Supervised Learning \and Multiple Instance Learning}
\end{abstract}
\vspace{-3 em}
\section{Introduction}
\vspace{-1 em}
%\jlreplace{
%    Pulmonary embolism (PE) is a condition caused due to artery blockage in the lung. CT pulmonary angiography (CTPA) is a standard medical imaging modality used for PE diagnosis~\cite{PEDiagnosis}. A single CTPA scan may have hundreds of scans that require a detailed review for accurate identification of PE. Also, multiple labels (image-level, exam-level, and informational) can be assigned to a single scan. These factors make the diagnosis time exhaustive and prone to over-diagnosis. Advancement in deep learning has led to the development of sophisticated computer-aided diagnosis tools that can address these challenges and expedite the diagnosis process~\cite{tajbakhsh2015computer,pmlr-v116-rajan20a,penet2020}. {}{However, different deep learning approaches can be applied to PE detection, and one method can significantly lead the other. In this work, we compare possible candidates through detailed experiments and infer an optimal performance approach. }
%}
{
    Pulmonary embolism (PE) represents a thrombus (occasionally colloquially, and incorrectly, referred to as a “blood clot”), usually originating from a lower extremity or pelvic vein, that travels to the blood vessels in the lung, causing vascular obstruction. PE causes more deaths than lung cancer, breast cancer, and colon cancer combined~\cite{USDHHS2008}. 
    %PE is a blockage in the pulmonary arteries, and most patients who succumb to PE do so within the first few hours following the event. %therefore, a key clinical challenge is to quickly and correctly diagnose patients with PE, so that hazardous yet life-saving therapy can be prescribed appropriately~\cite{torbicki2008task}. 
    % Diagnosing PE has two {\em separate} but {\em intertwined} objectives: patient-level diagnosis (quickly excluding non-PE patients and dispatching PE-patients to treatment) and embolus-level detection (localizing individual emboli and supporting per-sonalized medicine through risk stratification). 
    The current test of choice for PE diagnosis is CT pulmonary angiogram (CTPA)~\cite{stein2006multidetector}, but \iffalse Each CTPA scan contains hundreds of axial images, and interpreting these images is complex and time consuming.\fi studies have shown 14\% under-diagnosis and 10\% over-diagnosis with CTPA~\cite{lucassen2013concerns}. \iffalse To address this issue,\fi Computer-aided diagnosis (CAD) has shown great potential for improving the imaging diagnosis of PE~\cite{masutani2002computerized,liang2007computer,zhou2009computer,tajbakhsh2015computer,pmlr-v116-rajan20a,penet2020,zhou2019models,zhou2021towards,zhou2021active,zhou2017fine}. However, recent research in deep learning across academia and industry produced numerous architectures, various model initialization, and distinct learning paradigms, resulting in many competing approaches to CAD implementation in medical imaging producing great confusion in the CAD community. To address this confusion and develop an optimal approach, we wish  to answer a critical question: {\em What deep learning architectures, model initialization, and learning paradigms should be used for CAD applications in medical imaging?} To answer the question, we have conducted extensive experiments with various deep learning methods applicable for PE diagnosis at both image and exam levels using a publicly available PE dataset~\cite{RsnaDataset}.
}

\jlreplace{
    \jlreplace{Convolutional}{In terms of architectures, convolutional} neural networks (CNNs) have been the default architecture choice for the applications like classification and segmentation. CNNs have witnessed rapid development and have been successfully applied in the medical domain also~\cite{LITJENS2017,Deng2020}. Of late, transformer based architectures have been used as a substitute for CNNs~\cite{dosovitskiy2020vit,han2021transformer,touvron2020deit}.  Vision transformer (ViT)  is an adaptation of the transformer for images, lacks convolutional filters, and utilizes a multi-head self-attention mechanism~\cite{dosovitskiy2020vit,Vaswani2017Transformer}. However, irrespective of the architecture type, a large amount of data is required for optimal training. While many samples can be easily acquired, the challenge arises in the annotation process, especially in the medical domain. A counter to the data scarcity problem exists in the form of transfer learning~\cite{Liang2016}. In one heuristic of transfer learning, a pre-trained architecture is used for direct feature extraction on the target dataset. While in another approach, the target dataset is used for fine-tuning the pre-trained model. 
    
    Usually, the supervised learning paradigm is used for training a model, and hence annotations are required.  Self-supervised learning is a different approach that does not need explicit annotation, and features learned through a pretext task can be used in the target downstream task~\cite{jing2020}. 
%    In terms of the classification approaches, conventionally, each instance is assigned a single label. Another approach called multiple instance learning (MIL) makes a single prediction for a bag of instances, and hence, multiple samples {}{(a bag)} are assigned a single label~\cite{ITW2018}. MIL is helpful for predicting exam-level labels in PE diagnosis as only a single label is required for each exam {}{(a bag of images)}. In this work, we explore the utility of above discussed approaches on a publicly available PE dataset~\cite{RsnaDataset}. An analysis is provided for image-level predictions as well as exam-level predictions. Furthermore, we compare and contrast CNN and ViT concerning training from scratch and transfer learning. A comparison between supervised learning and self-supervised learning for image-level predictions is also provided. Finally, for exam-level predictions,  we compare the performance of conventional classification and MIL. 
}

    \iffalse In terms of architectures,\fi Convolutional neural networks (CNNs) have been the default architectural choice for classification and segmentation in medical imaging~\cite{LITJENS2017,Deng2020}. Nevertheless, transformers have proven to be powerful in Natural Language Processing (NLP)~\cite{devlin2018bert,brown2020language}, and have been quickly adopted for image analysis~\cite{dosovitskiy2020vit,han2021transformer,touvron2020deit}, leading to vision transformer (ViT)\iffalse, which utilizes multi-head self-attention and involves no convolution\fi~\cite{dosovitskiy2020vit,Vaswani2017Transformer}. Therefore, to assess architecture performance, we compared ViT with {10} CNNs variants for classifying PE. Regardless of the architecture, training deep models generally requires massive carefully labeled training datasets~\cite{haghighi2021transferable}. However, \iffalse in medical imaging, \fi it is often prohibitive to create such large annotated datasets in medical imaging; therefore, fine-tuning models from ImageNet has become the \textit{de facto} standard~\cite{tajbakhsh2016convolutional,shin2016deep}. As a result, we benchmarked various models pre-trained on ImageNet against training from scratch. 
    
    Supervised learning is currently the dominant approach for classification and segmentation in medical imaging, which offers expert-level and sometimes even super-expert-level performance. Self-supervised learning (SSL) has recently garnered attention for its capacity to learn generalizable representations without requiring expert annotation~\cite{jing2020,haghighi2020learning}. The idea is to pre-train models on pretext tasks and then fine-tune the pre-trained models to the target tasks. We evaluated {14} different SSL methods for PE diagnosis. In contrast to conventional classification (CC), which predicts a label for each instance, multiple instance learning (MIL) makes a single prediction for a bag of instances; that is, multiple instances belonging to the same ``bag'' are assigned a single label~\cite{ITW2018}. MIL is label efficient because only a single label is required for each exam (an exam is therefore a ``bag'' of instances). Therefore, it is important to ascertain the effectiveness of MIL for PE diagnosis at the exam level.
    
    %In this work, we explore the utility of above discussed approaches on a publicly available PE dataset~\cite{RsnaDataset}. An analysis is provided for image-level predictions as well as exam-level predictions. Furthermore, we compare and contrast CNN and ViT concerning training from scratch and transfer learning. A comparison between supervised learning and self-supervised learning for image-level predictions is also provided. Finally, for exam-level predictions,  we compare the performance of conventional classification and MIL.    
{In summary, our work offers three contributions: (1) a comprehensive analysis of competitive deep learning methods for PE diagnosis; \iffalse at both image and exam levels using CT scans;\fi (2) extensive experiments that compare \iffalse deep learning methods for PE diagnosis in terms of \fi architectures, \iffalse (CNNs vs transformers),\fi model initialization, \iffalse (supervised vs. self-supervised learning; conventional classification vs. multiple instance learning),\fi and learning paradigms; \iffalse (fine-tuning vs. training from scratch);\fi and (3) an optimal approach for detecting PE, achieving an AUC gain of 0.2\% and 1.05\% at the image and exam levels, respectively, compared with the state-of-the-art performance.
}
\vspace{-1 em}
\section{Materials}
\vspace{-1 em}
The Radiological Society of North America (RSNA) Pulmonary Embolism Detection Challenge (RSPED) aims to advance computer-aided diagnosis for pulmonary embolism detection~\cite{RsnaChallenge}. The dataset consists of 7,279 CTPA exams, with a varying number of images in each exam, using an image size of $512 \times 512$ pixels. The {test} set is created by randomly sampling 1000 exams, and the remaining 6279 exams form the training set. Correspondingly, there are  1,542,144 and 248,480 images in the training and {test} sets, respectively. This dataset is annotated at both image and exam level; that is, each image has been annotated as either PE presence or PE absence. Each exam has been annotated for an additional nine labels (see Table~\ref{pePatientRes}).

\begin{figure}[!t]
\centering
%\centerline{
\subfloat[]{
\includegraphics[height=1.7 cm, width=1.7cm]{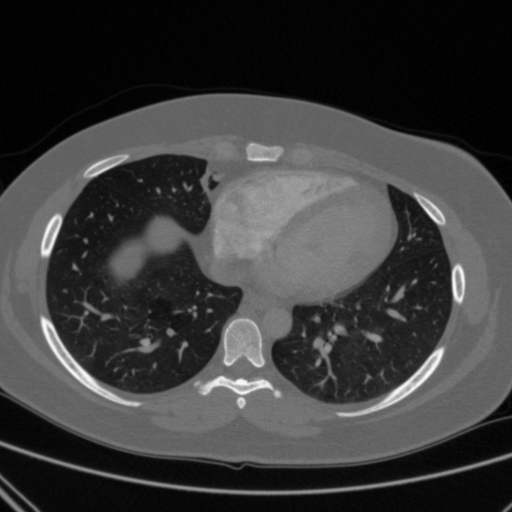}
\label{fig:subfig1}}
\subfloat[]{
\includegraphics[height=1.7 cm, width=1.7cm]{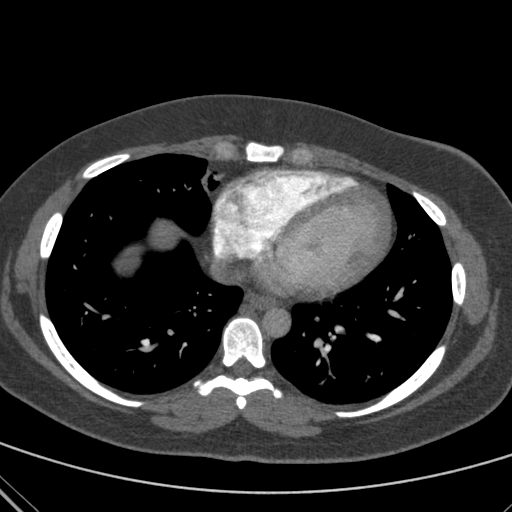}
\label{fig:subfig2}}
\subfloat[]{
\includegraphics[height=1.7 cm, width=1.7cm]{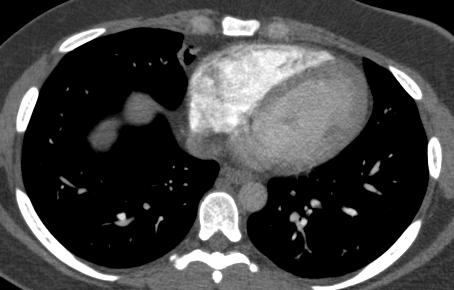}
\label{fig:subfig3}}
\qquad
\subfloat[]{
\includegraphics[height=1.7 cm, width=1.7cm]{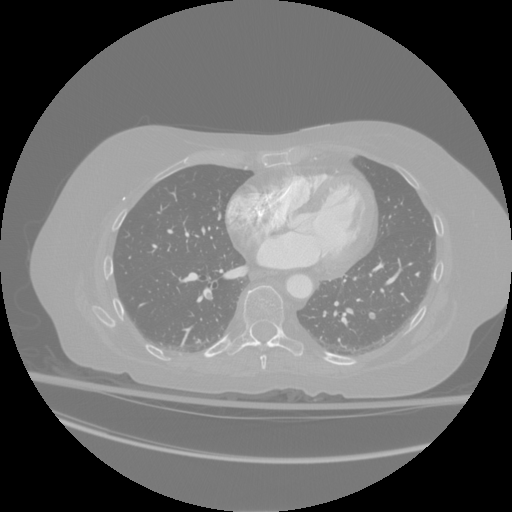}
\label{fig:subfig1}}
\subfloat[]{
\includegraphics[height=1.7 cm, width=1.7cm]{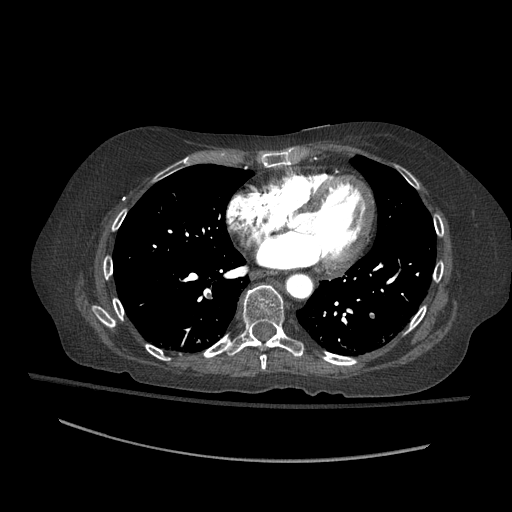}
\label{fig:subfig2}}
\subfloat[]{
\includegraphics[height=1.7 cm, width=1.7cm]{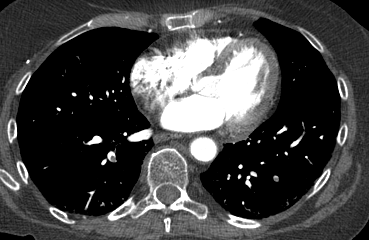}
\label{fig:subfig3}}
\caption{\small The pre-processing steps for image-level classification. (a,d) {original CT images},  (b,e) after windowing, and (c,f) after lung localization. For windowing, pixels above 450 HU and below -250 HU were clipped to 450 HU and -250 HU, respectively.}

\label{preprocessing}
\vspace{-1 em}
\end{figure}

%\subsection{Implementation}
Similar to the first place solution for this challenge, lung localization and windowing have been used as pre-processing steps \cite{FirstPlaceSolution} . Lung localization removes the irrelevant tissues and keeps the region of interest in the images, whereas windowing highlights the pixel intensities within the range of [100, 700]. Also, the images are resized to $576 \times 576$ pixels. \figurename~\ref{preprocessing} illustrates these pre-processing steps in detail. We considered three adjacent images from an exam as the 3-channel input of the model.

\vspace{-1 em}
\section{Methods}
\vspace{-0.5 em}
\subsection{Image-level Classification}
\vspace{-0.2 em}
Image-level classification refers to determining the presence or absence of PE for each image. 
In this section, we describe the configurations of supervised and self-supervised transfer learning in our work.
\vspace{-1 em}
\paragraph{\textnormal{\textbf{Supervised Learning:}}}
 The idea is to pre-train models on ImageNet with ground truth and then fine-tune the pre-trained models for PE diagnosis, in which we examine ten different CNN architectures (\figurename~\ref{fig:backbone_sub1}). Inspired by SeResNext50 and SeResNet50, we introduced squeeze and excitation (SE) block to the Xception architecture (SeXception). These CNN architectures were pre-trained on ImageNet\footnote{We pre-trained SeXception on ImageNet and the others are taken from \href{https://github.com/Cadene/pretrained-models.pytorch}{PyTorch}}. We also explored the usefulness of vision transformer (ViT), where the images are reshaped into a sequence of patches. We experimented with ViT-B\_32 and ViT-B\_16 utilizing $32 \times 32$ and $16 \times 16$ patches, respectively~\cite{bert2019}. Again, ViT architectures were pre-trained on ImageNet21k. Upscaling the image for a given patch size will effectively increase the number of patches, thereby enlarging the size of the training dataset; models are also trained on different sized images. Similarly, the number of patches increases with a decrease in the patch size.
\paragraph{\textnormal{\textbf{Self-Supervised Learning (SSL):}}}
In self-supervised transfer learning, the model is pre-trained on  ImageNet without ground truth and then fine-tuned for PE diagnosis. Self-supervised learning has gained recent attention~\cite{sela,caron2018deep,zbontar2021barlow}. With the assistance of strong augmentation and comparing \iffalse them with \fi different contrastive losses, a model can learn meaningful information~\cite{hu2021well} without annotations. These architectures are first trained for a pretext task; for example, reconstructing the original image from its distorted version. Then the models are fine-tuned for a different task, in our case, PE detection. We pre-train models through {14} different SSL approaches, all of which used ResNet50 as the backbone.

% for Nahid -> check the 14 summarized SSL paper (from the introduction): With the \drmgreplace{help}{assistance} of strong augmentation and comparing \iffalse them with \fi different contrastive losses, a model can learn meaningful information~\cite{hu2021well} without annotations. These architectures are first trained for a pretext task, for example, reconstructing the original image from its distorted version.
\vspace{-1 em}
\subsection{Exam-level Classification}
\vspace{-0.5 em}
Apart from the image-level classification, the RSPED dataset also provides exam-level labels, in which only one label is assigned for each exam. For this task, we used the features extracted from the models trained for image-level PE classification, and explored two learning paradigms as follows:
\vspace{-0.5 em}
\paragraph{\textnormal{\textbf{Conventional Classification (CC):}}}
\label{cc}
We stacked all the extracted features together resulting in an $N \times M$ feature for each exam, where $N$ and $M$ denote the number of images per exam and the dimension of the image feature, respectively. However, as $N$ varies from exam to exam, the feature was reshaped to $K \times M$. Following ~\cite{FirstPlaceSolution}, we set the $K$ equal to 192 in the experiment. The features were then fed into a bidirectional Gated Recurrent Unit (GRU) followed by pooling and fully connected layers to predict exam-level labels. 
\vspace{-0.5 em}
\paragraph{\textnormal{\textbf{Multiple Instance Learning (MIL):}}}
MIL is annotation efficient as it does not require annotation for each instance \cite{mil-review}. An essential requirement for MIL is permutation invariant MIL pooling~\cite{ITW2018}. Both max operators and attention-based operator are used as MIL pooling ~\cite{ITW2018}, and we experimented with a combination of these approaches. The MIL approach is innate for handling varying images ($N$) in the exams and does not require any reshaping operation as does Conventional Classification (CC). For MIL, we exploited the same architecture as in CC by replacing pooling with MIL pooling~\cite{ITW2018}. 
\iffalse
\drmgreplace{RSNA}{The Radiological Society of North America (RSNA) } Pulmonary Embolism Detection Challenge (RSPED) was organized by the RSNA to foster machine-learning algorithms \drmgadd{development} for pulmonary embolism detection~\cite{RsnaChallenge}. The dataset consists of 7279 \drmgadd{CTPA} exams and it is annotated at the image level as well as the exam level. Each exam has a varying number of CT images with the size of $512 \time 512$. We split the dataset into training (6,279 exams; 1,542,144 images) and testing (1,000 exams; 248,480 images) sets at exam level. As because the challenge has ended and they did not make the official testing set publicly available, we used our split for the experiments. 
\fi
\vspace{-0.8 em}
\section{Results and Discussion}
\vspace{-0.5 em}
\iffalse
Results & Discussion
Figure 3: https://raphaelvallat.com/correlation.html
In this section, we discuss the results and observations based on the experiments discussed in the previous section. \fi

\begin{figure}[!t]
\centerline{
    \subfloat[]{
    \includegraphics[height=4.25cm, width=8.10cm]{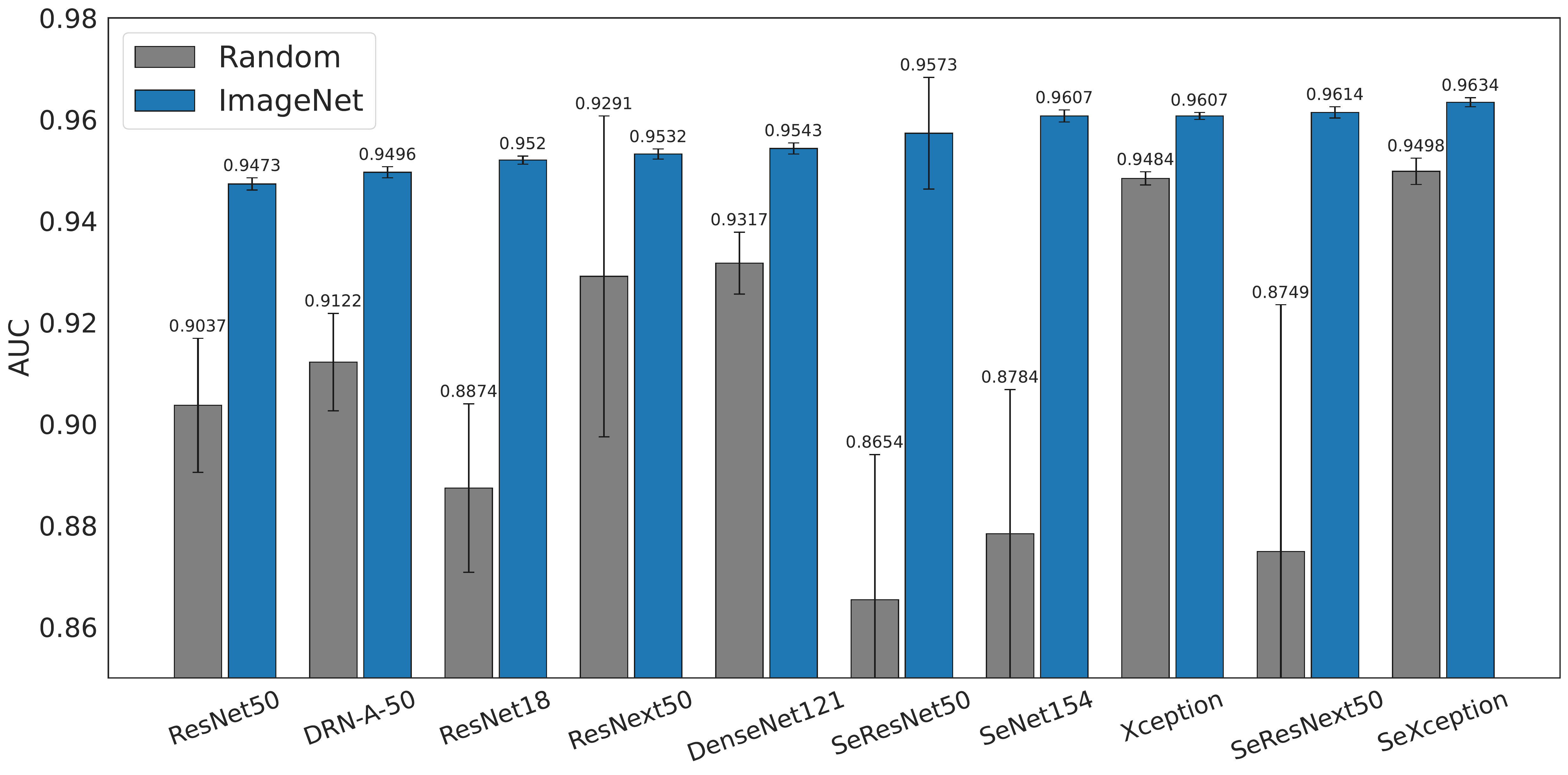}
    \label{fig:backbone_sub1}}
    \subfloat[]{
    \includegraphics[height=4cm, width=3.85cm ]{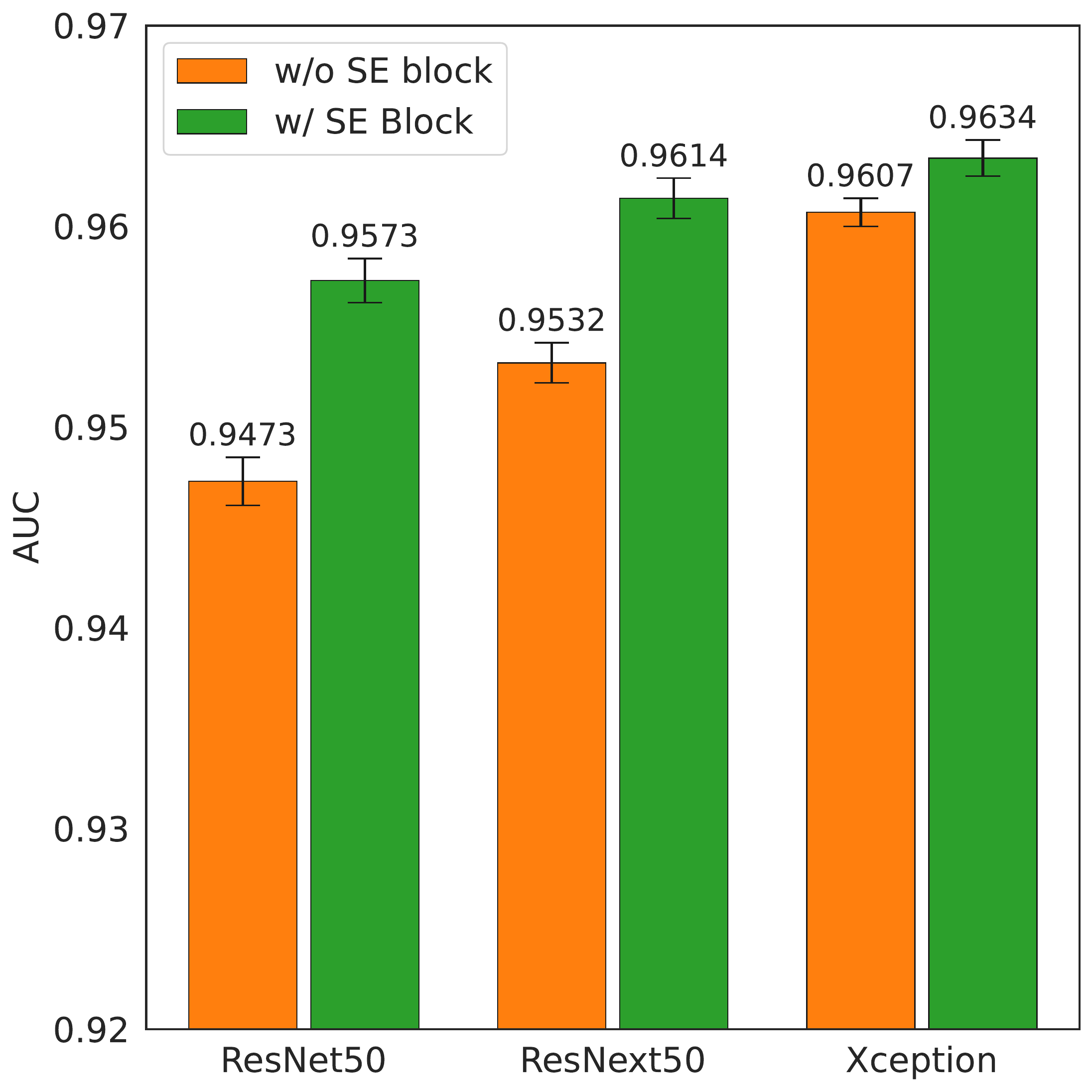}
    \label{fig:backbone_sub2}}}
    \caption{{}{(a) For all 10 architectures, transfer learning outperformed random ini in image-level PE classification}, in spite of the pronounced difference between ImageNet and RSPED. Mean AUC and standard deviation over ten runs are reported for each architecture. Compared with the previous state of the art (SeResNext50), the SeXception architecture achieved a significant improvement ($p$ = 1.68E-4). (b) We observed a performance gain with the help of SE block. Note that, all the architectures under comparison were pre-trained from ImageNet.} \label{fig:backbone_sub1}
    \vspace{-2 em}
\end{figure}
\begin{figure}[!t]
    \includegraphics[width=0.8\textwidth]{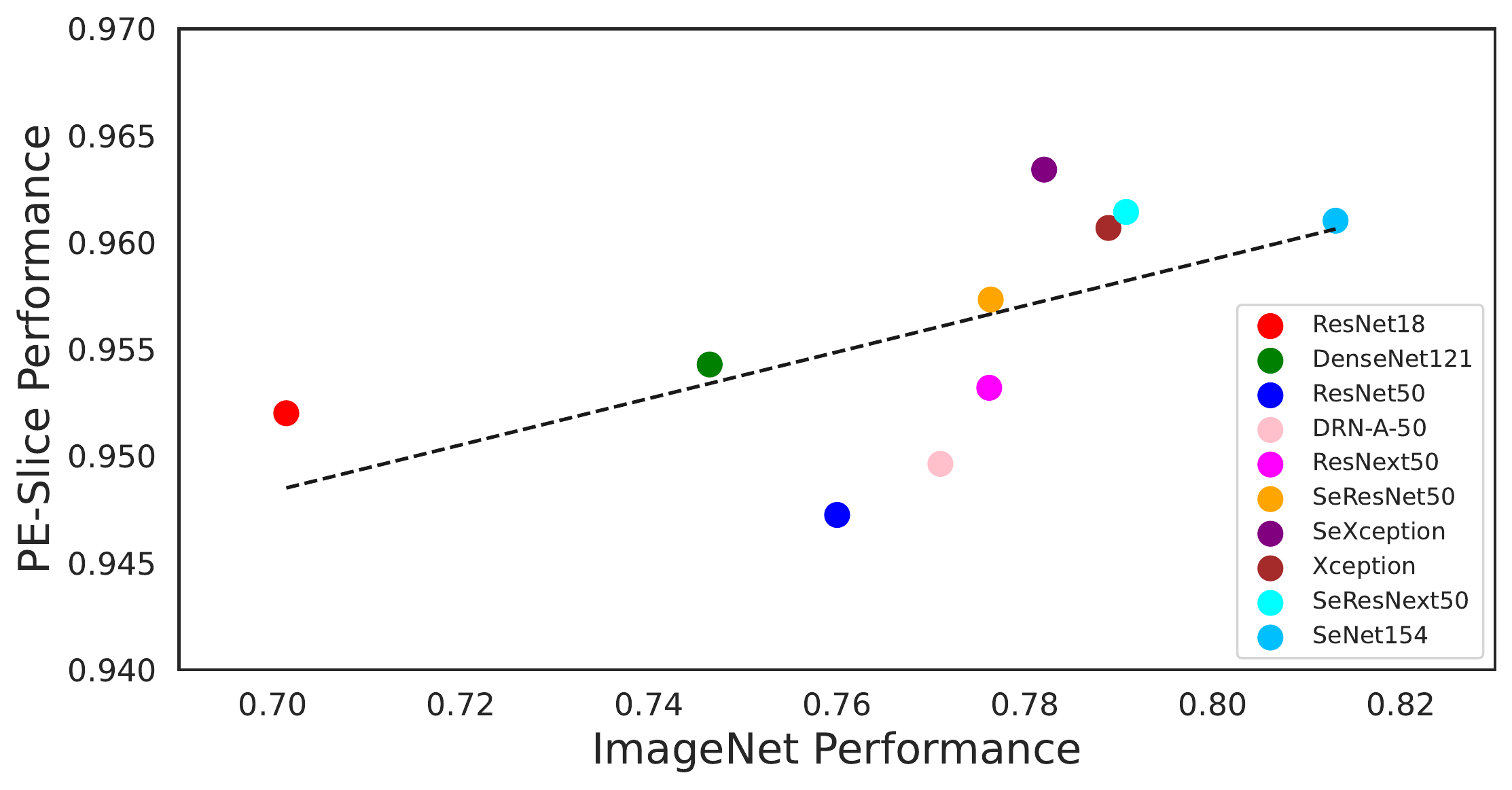}\centering
    \caption{{}{There was a positive correlation between the results on ImageNet and RSPED ($R = 0.5914$), suggesting that the transfer learning performance could be inferred by ImageNet pre-training performance.}} \label{imagenet_pe_corr}
     \vspace{-2 em}
\end{figure}
\begin{figure}[!t]
\includegraphics[width=0.6\textwidth]{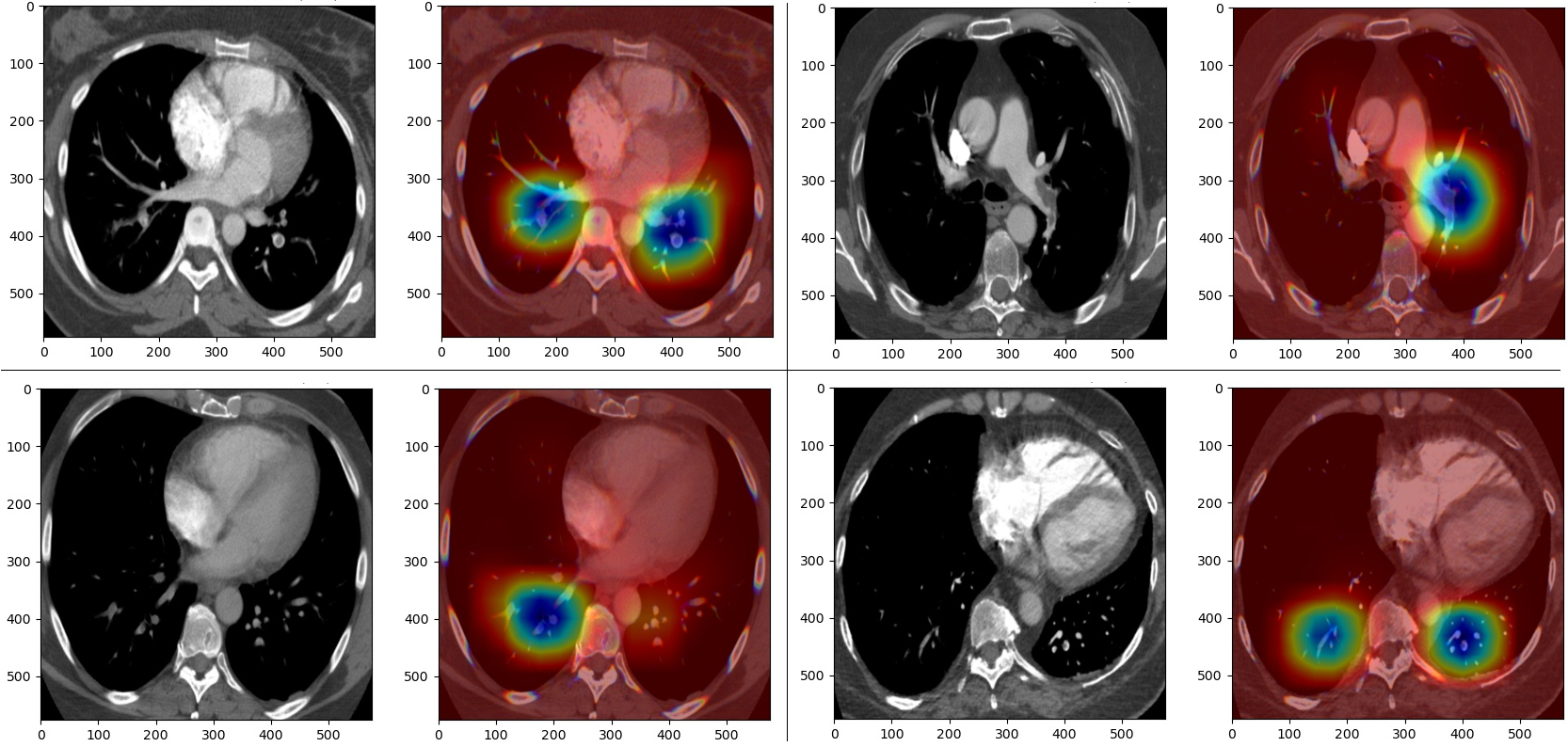}\centering
\caption{{}{The SeXception attention map highlighted the potential PE location in the image using GradCam++.}} \label{gradcam}
\end{figure}

\vspace{0.2 em}
\paragraph{\textnormal{\textbf{1. Transfer learning significantly improves the performance of image-level classification despite the modality difference between the source and target datasets.}}}
\figurename~\ref{fig:backbone_sub1} shows a significant performance gain for every pre-trained model compared with random initialization. There is also a positive correlation of 0.5914 between ImageNet performance and PE classification performance across different architectures (\figurename~\ref{imagenet_pe_corr}), indicating that useful weights learned from ImageNet can be successfully transferred to the PE classification task, despite the modality difference between the two datasets. With the help of GradCam++~\cite{pygradcam_pp}, we also visualized the attention map of SeXception, the best performing architecture. As shown in Fig~\ref{gradcam}, the attention map can successfully highlight the potential PE location in the image.
\vspace{-0.5 em}
\paragraph{\textnormal{\textbf{2. Squeeze and excitation (SE) block enhances  CNN performance.}}} Despite fewer parameters compared with many other architectures, SeXception provides an optimal average AUC of 0.9634. SE block enables an architecture to extract informative features by fusing spatial and channel-wise information~\cite{seblock}. Thus, the SE block has led to performance improvements from ResNet50 to SeResNet50, ResNext50 to SeResNext50 and from Xception to SeXception (\figurename~\ref{fig:backbone_sub2}).
\begin{figure}[!t]
\centerline{
\includegraphics[width=0.98\textwidth]{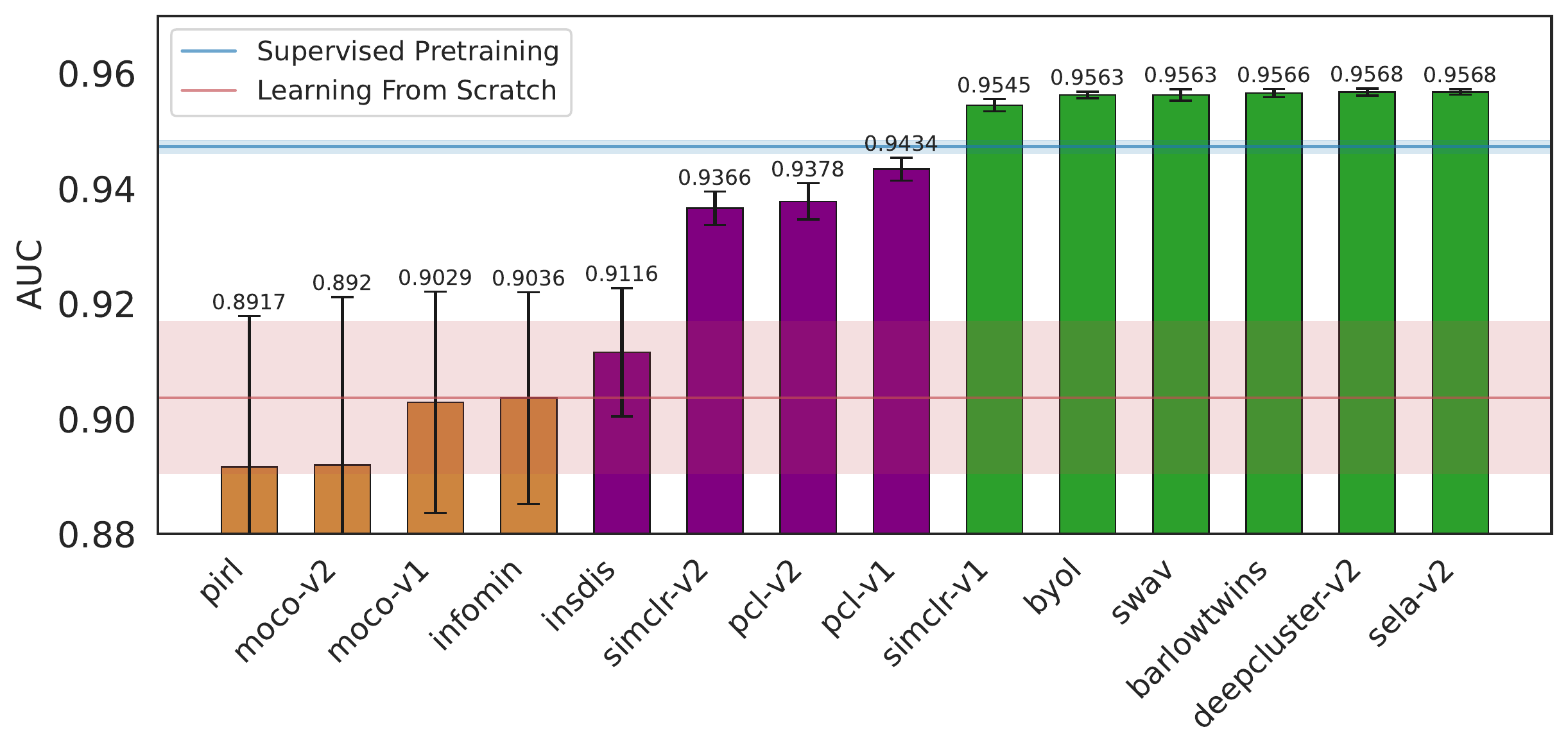}}\centering
\caption{ {}{Self-supervised pre-training extracted more transferable features compared with supervised pre-training. The blue and red dashed lines represent supervised pre-training and learning from scratch with standard deviation (shaded), respectively. 6 out of 14  SSL methods (in green) outperformed the supervised pre-training. All the reported methods had ResNet50 as the backbone.}} \label{sslFT}
\vspace{-2 em}
\end{figure}
\paragraph{\textnormal{\textbf{3. Transfer learning with a self-supervised paradigm produces better results than its supervised counterparts.}}} As summarized in \figurename~\ref{sslFT}, SeLa-v2~\cite{sela} and DeepCluster-v2~\cite{caron2018deep} achieved the best AUC of 0.9568, followed by Barlow Twins~\cite{zbontar2021barlow}. Six out of fourteen SSL models performed better than supervised pre-trained ResNet50 (\figurename~\ref{sslFT}). Further experiments can be conducted with other backbones (\figurename~\ref{fig:backbone_sub1}) to explore if SSL models outperform other supervised counterparts as well, a subject of future work.
\begin{table}[!h]
\caption{ViT performs inferiorly compared with CNN for image-level PE classification. For both architectures (ViT-B\_32 and ViT-B\_16), random initialization provides the worst performance. Both increasing the image size and reducing the patch size can enlarge the training set and therefore lead to an improved performance. Finally, similar to CNNs, initializing ViTs on ImageNet21k provided significant performance gain, indicating the usefulness of  transfer learning.}

\label{vit_results}
% \centerline{
% \resizebox{0.8\textwidth}{!}{%
% \begin{tabular}{p{0.15\linewidth}P{0.15\linewidth}P{0.15\linewidth}P{0.15\linewidth}P{0.15\linewidth}}
% \hline
% \multicolumn{5}{c}{PE AUC with vision transformer (ViT)}       \\ \hline
% Model     & Image Size & Patch Size & Initialization & Val AUC \\ \hline
% SeXception & 576       & NA        & ImageNet        & \textbf{0.9634}  \\ \hline
% ViT-B\_32 & 512        & 32         & Random         & 0.8212  \\ \
% ViT-B\_32 & 224        & 32         & ImageNet21k    & 0.8456  \\ 
% ViT-B\_32 & 512        & 32         & ImageNet21k    & 0.8847  \\ \hline
% ViT-B\_16 & 512        & 16         & Random         & 0.8385  \\ 
% ViT-B\_16 & 224        & 16         & ImageNet21k    & 0.8826  \\ 
% ViT-B\_16 & 512        & 16         & ImageNet21k    & 0.9065  \\ 
% ViT-B\_16 & 576        & 16         & ImageNet21k    & \emph{0.9179}  \\ \hline
% \end{tabular}%
% }}
\centerline{
\resizebox{1.0\textwidth}{!}{%
\begin{tabular}{p{0.2\linewidth}P{0.2\linewidth}P{0.2\linewidth}P{0.3\linewidth}P{0.2\linewidth}}
\shline
\multicolumn{5}{c}{PE AUC with vision transformer (ViT)}       \\ \hline
Model     & Image Size & Patch Size & Initialization & Val AUC \\ \hline
SeXception & 576       & NA        & ImageNet        & \textbf{0.9634}  \\ \hline
ViT-B\_32 & 512        & 32         & Random         & 0.8212  \\
ViT-B\_32 & 224        & 32         & ImageNet21k    & 0.8456  \\ 
ViT-B\_32 & 512        & 32         & ImageNet21k    & 0.8847  \\ \hline
ViT-B\_16 & 512        & 16         & Random         & 0.8385  \\ 
ViT-B\_16 & 224        & 16         & ImageNet21k    & 0.8826  \\ 
ViT-B\_16 & 512        & 16         & ImageNet21k    & 0.9065  \\ 
ViT-B\_16 & 576        & 16         & ImageNet21k    & \emph{0.9179}  \\ \shline
% \vspace{-2 em}
\end{tabular}%
}}
\vspace{-1.2 em}
\end{table}
\begin{table}[!h]
        \caption{{}{The features extracted by the models trained for image-level classification were helpful for exam-level classification. However, no model performed consistently best for all labels. We report the mean AUC over 10 runs and bold the optimal results for each label. The Xception architecture achieved a significant improvement ($p$ = 5.34E-12) against the previous state of the art$^{\dagger}$.}}\label{pePatientRes}
        \centerline{
            \resizebox{ 1.0 \textwidth}{!}{%
            \begin{tabular}{p{0.25\linewidth}P{0.14\linewidth}P{0.14\linewidth}P{0.14\linewidth}P{0.14\linewidth}P{0.14\linewidth}P{0.14\linewidth}}
            \shline
            Labels &  SeResNext50$^{\dagger}$ & Xception & SeXception & DenseNet121 & ResNet18 & ResNet50\\
            \hline
            NegExam PE &  0.9137 & 0.9242 & \textbf{0.9261} & 0.9168 & 0.9141 & 0.9061\\
            Indetermine &  0.8802 & 0.9168 & 0.8857 & 0.9233 & 0.9014 & \textbf{0.9278}\\
            Left PE & 0.9030 & 0.9119 & 0.9100 & \textbf{0.9120} & 0.9000 & 0.8965\\
            Right PE & 0.9368 & 0.9419 & \textbf{0.9455} & 0.9380 & 0.9303 & 0.9254\\
            Central PE & 0.9543 & 0.9500 & 0.9487 & \textbf{0.9549} & 0.9445 & 0.9274\\
            RV LV Ratio$\geq$1 & 0.8902 & \textbf{0.8924} & 0.8901 & 0.8804 & 0.8682 & 0.8471\\
            RV LV Ratio$<$1 & 0.8630 & 0.8722 & \textbf{0.8771} & 0.8708 & 0.8688 & 0.8719\\
            Chronic PE & 0.7254 & \textbf{0.7763} & 0.7361 & 0.7460 & 0.6995 & 0.6810\\
            Acute\&Chronic PE & \textbf{0.8598} & 0.8352 & 0.8473 & 0.8492 & 0.8287 & 0.8398\\
            \hline
            Mean AUC & 0.8807 & \textbf{0.8912} & 0.8852 & 0.8879 & 0.8728 & 0.8692\\            
            \shline
            \end{tabular}
            }}
            \vspace{- 2 em}
\end{table}
\vspace{- 1 em}
\paragraph{\textnormal{\textbf{4. CNNs have better performance than ViTs.}}}
{As shown in} \tablename~\ref{vit_results}, random initialization provides a significantly lower performance than ImageNet pre-training. The best AUC of $0.9179$ is obtained by ViT-B\_16 with image size $576 \times 576$ and ImageNet21k initialization. However, this performance is inferior to the optimal CNN architecture (SeXception) by a significant margin of approximately 4\%. We attribute this result to the absence of convolutional filters in ViTs.
\vspace{-0.7 em}
\paragraph{\textnormal{\textbf{5. Conventional classification (CC) marginally outperforms MIL.}}} The results of CC for exam-level predictions are summarized in Table \ref{pePatientRes}. Although SeXception performed optimally for image-level classification (see ~\figurename~\ref{fig:backbone_sub1}), the same is not true for exam-level classification. There is no architecture that performs optimally across all labels, but overall, Xception shows the best AUC across nine labels. The results of MIL for exam-level predictions are summarized in Table \ref{mil_auc}.
Xception achieved the best AUC with a combination of attention and max pooling. Similar to CC approach, no single MIL method performs optimally for all labels. However, Xception shows the best mean AUC of $0.8859$ with Attention and Max Pooling across all labels. \textit{Furthermore, the AUC for MIL is marginally lower than CC ($0.8859$ vs. $0.8912$) but the later requires additional prepossessing steps (\ref{cc}). More importantly, MIL provides a more flexible approach and can easily handle varying number of images per exam.}
Based on result \#3, the performance of exam-level classification may be improved by incorporating the features from SSL methods.
% Shiv -> bold best for each row
% Shiv -> but the later requires additional prepossessing steps. --> what preprocessing step? is it the 192 value? Need to reference the explanation of the preprocessing steps.
\vspace{-0.4 em}
\begin{table}[!t]
\caption{The performance varies with pooling strategies for MIL. Attention and Max Pooling (AMP) combines the output of Max Pooling (MP) and Attention Pooling (AP). MIL utilized the feature extracted by the model trained for image-level PE classification. For all three architectures, the best mean AUC is obtained by AMP, highlighting the importance of combining AP and MP.}
\label{mil_auc}
\centerline{
\resizebox{1.0\textwidth}{!}{%
% \begin{tabular}{cccc|ccc|ccc}
\begin{tabular}{p{0.25\linewidth}P{0.10\linewidth}P{0.10\linewidth}P{0.10\linewidth}|P{0.10\linewidth}P{0.10\linewidth}P{0.10\linewidth}|P{0.10\linewidth}P{0.10\linewidth}P{0.10\linewidth}}
\shline
Architecture               & \multicolumn{3}{c|}{SeResNeXT50} & \multicolumn{3}{c|}{Xception}      & \multicolumn{3}{c}{SeXception} \\ \hline
Labels/Pooling             & AMP       & MP       & AP       & AMP             & MP     & AP     & AMP       & MP       & AP       \\ \hline
NegExam PE                 & 0.9138    & 0.9137   & 0.9188   & \textbf{0.9202}          & \textbf{0.9202} & 0.9172 & 0.9183    & 0.9137   & 0.9201   \\ 
Indetermine                & \textbf{0.9144}    & 0.9064   & 0.8986   & 0.8793          & 0.8580  & 0.8933 & 0.8616    & 0.8564   & 0.8499   \\ 
Left PE                    & \textbf{0.9122}    & 0.9059   & 0.9086   & 0.9106          & 0.9100   & 0.9032 & 0.9042    & 0.9004   & 0.9024   \\ 
Right PE                   & 0.9340     & 0.9345   & 0.9373   & 0.9397          & 0.9366 & 0.9397 & 0.9403    & 0.9383   & \textbf{0.9412}   \\ 
Central PE                 & \textbf{0.9561}    & 0.9537   & 0.9529   & 0.9487          & 0.9465 & 0.9507 & 0.9472    & 0.9424   & 0.9453   \\ 
RV LV Ratio$\geq$1 & 0.8813    & 0.8774   & 0.8822   & \textbf{0.8920}           & 0.8871 & 0.8819 & 0.8827    & 0.8779   & 0.8813   \\ 
RV LV Ratio$<$1    & 0.8597    & 0.8606   & 0.862    & \textbf{0.8644}          & 0.8619 & 0.8567 & 0.8676    & 0.8642   & \textbf{0.8644}   \\ 
Chronic PE                 & 0.7304    & 0.7256   & 0.7233   & \textbf{0.7788}          & 0.7664 & 0.7699 & 0.7334    & 0.7168   & 0.7342   \\ 
Acute\&Chronic PE          & \textbf{0.8453}    & 0.8470    & 0.8228   & 0.8392          & 0.8396 & 0.8350  & 0.8405    & 0.8341   & 0.8367   \\ \hline
Mean AUC                   & 0.8830     & 0.8805   & 0.8785   & \textbf{0.8859} & 0.8807 & 0.8831 & 0.8773    & 0.8716   & 0.8751   \\ \shline
%Weighted Loss              & 0.1883    & 0.1899   & 0.1885   & \textbf{0.1866} & 0.1886 & 0.1907 & 0.1892    & 0.1936   & 0.1903   \\ \hline
\end{tabular}%
}}
\vspace{- 0.7 em}
\end{table}
\vspace{-0.7 em}
\paragraph{\textnormal{\textbf{{}{6. The optimal approach:}}}}The existing first place solution \cite{FirstPlaceSolution} utilizes SeResNext50 for image-level and CC for exam-level classification. Compared with their solution, our optimal approach achieved an AUC gain of 0.2\% and 1.05\% for image-level and exam-level PE classification, respectively. Based on our rigorous analysis, the optimal architectures for the tasks of image-level and exam-level classification are SeXception and Xception.
\vspace{-1 em}
\section{Conclusion}
\vspace{-1 em}
We analyzed different deep learning architectures, model initialization, and learning paradigms for image-level and exam-level PE classification on CTPA scans. We benchmarked CNNs, ViTs, transfer learning, supervised learning, SSL, CC, and MIL, and concluded that transfer learning and CNNs are superior to random initialization and ViTs. Furthermore, SeXception is the optimal architecture for image-level classification, whereas Xception performs best for exam-level classification. A detailed study of SSL methods will be undertaken in future work.

% We have tried to improve the performance for patient-level classification by using different methods but our analysis favours the bidirectional-GRU approach used by the first place solution.

\smallskip
\noindent\textbf{Acknowledgments:}
This research has been supported partially by ASU and Mayo Clinic through a Seed Grant and an Innovation Grant, and partially by the NIH under Award Number R01HL128785.  The content is solely the responsibility of the authors and does not necessarily represent the official views of the NIH. This work has utilized the GPUs provided partially by the ASU Research Computing and partially by the Extreme Science and Engineering Discovery Environment (XSEDE) funded by the National Science Foundation (NSF) under grant number ACI-1548562. We thank Ruibin Feng for helping us some experiments. The content of this paper is covered by patents pending.

% Nahid -> Comment this line
% \newpage
\bibliographystyle{unsrt}
\bibliography{main}

% \bibitem{rsna_1}
% E Colak, FC Kitamura, SB Hobbs, et al. The RSNA Pulmonary Embolism CT Dataset [https://pubs.rsna.org/doi/full/10.1148/ryai.2021200254]. Radiology: Artificial Intelligence 2021;3:2.

% \bibitem{pygradcam_pp}
% Jacob Gildenblat and contributors. PyTorch library for CAM methods. https://github.com/jacobgil/pytorch-grad-cam, 2021.

% \bibitem{seblock}
% Hu, J., Shen, L., \& Sun, G. (2018). Squeeze-and-excitation networks. In Proceedings of the IEEE conference on computer vision and pattern recognition (pp. 7132-7141).

% \bibitem{full_fine}
% Tajbakhsh, N., Shin, J. Y., Gurudu, S. R., Hurst, R. T., Kendall, C. B., Gotway, M. B., & Liang, J. (2016). Convolutional neural networks for medical image analysis: Full training or fine tuning?. IEEE transactions on medical imaging, 35(5), 1299-1312.

% \bibitem{deepconv}
% C. Szegedy et al., Going deeper with convolutions ArXiv, 2014 [On- line]. Available: arXiv:1409.4842, to be published

% \end{thebibliography}

\newpage
\title{Supplementary Material}
% \maketitle 
\appendix

\section*{Appendix} 

% \section{Dataset}

% \section{Implementation}

\section{Tabular Results}
\begin{table}[h]
\begin{center}
\caption{
The tabular results of \figureautorefname~\ref{fig:backbone_sub1}.
% The comparison between random initialization and ImageNet pre-training for PE image-level classification, representing the tabular results of \figureautorefname~\ref{fig:backbone_sub1}. The best architecture is bolded. 
}
    % \resizebox{1.0\textwidth}{!}{%
    \begin{tabular}{p{0.25\linewidth}P{0.22\linewidth}P{0.22\linewidth}P{0.22\linewidth}}
    \shline
    Backbone & Parameters & Random & ImageNet\\
    \hline
    ResNet50 & 23,510,081 & 0.9037{\fontsize{6}{6}$\pm$}0.0132 & 0.9473{\fontsize{6}{6}$\pm$}0.0012\\
    DRN-A-50 & 23,510,081 & 0.9122{\fontsize{6}{6}$\pm$}0.0096 & 0.9496{\fontsize{6}{6}$\pm$}0.0011\\
    ResNet18 & 11,177,025 & 0.8874{\fontsize{6}{6}$\pm$}0.0166 & 0.9520{\fontsize{6}{6}$\pm$}0.0007\\
    DenseNet121 & 6,954,881 & 0.9317{\fontsize{6}{6}$\pm$}0.0060 & 0.9543{\fontsize{6}{6}$\pm$}0.0011\\
    SeResNet50 & 28,090,073 & 0.8654{\fontsize{6}{6}$\pm$}0.0293 & 0.9573{\fontsize{6}{6}$\pm$}0.0013\\
    SeNet154 & 27,561,945 & 0.8784{\fontsize{6}{6}$\pm$}0.0292 & 0.9607{\fontsize{6}{6}$\pm$}0.0012\\
    Xception & 20,809,001 & 0.9484{\fontsize{6}{6}$\pm$}0.0013 & 0.9607{\fontsize{6}{6}$\pm$}0.0007\\
    SeResNext50 & 27,561,945 & 0.8746{\fontsize{6}{6}$\pm$}0.0486 & 0.9614{\fontsize{6}{6}$\pm$}0.0011\\
    SeXception & 21,548,446 & \textbf{0.9498{\fontsize{6}{6}$\pm$}0.0025} & \textbf{0.9634{\fontsize{6}{6}$\pm$}0.0009}\\
    \shline
    \end{tabular}
    
\end{center}
\end{table}

\begin{table}[h]
\begin{center}
\caption{
The tabular results of \figureautorefname~\ref{sslFT}. 
% Experimental results of transfer learning from 14 self-supervised ImageNet pre-trained models for PE image-level classification, representing the tabular results of \figureautorefname~\ref{sslFT}. The best model is bolded. 6 out of 14 self-supervised models have outperformed supervised ResNet50 model.
}
    % \resizebox{.5\textwidth}{!}{%
    \begin{tabular}{p{0.25\linewidth}P{0.22\linewidth}}
    \shline
    Pre-training & Mean AUC\\
    \hline
    Random & 0.9037{\fontsize{6}{6}$\pm$}0.0132\\
    ImageNet & 0.9473{\fontsize{6}{6}$\pm$}0.0012\\
    \hline
    pirl & 0.8917{\fontsize{6}{6}$\pm$}0.0261\\
    moco-v2 & 0.8920{\fontsize{6}{6}$\pm$}0.0292\\
    moco-v1 & 0.9029{\fontsize{6}{6}$\pm$}0.0192\\
    infomin & 0.9036{\fontsize{6}{6}$\pm$}0.0184\\
    insdis & 0.9116{\fontsize{6}{6}$\pm$}0.0111\\
    simclr-v2 & 0.9366{\fontsize{6}{6}$\pm$}0.0029\\
    pcl-v2 & 0.9378{\fontsize{6}{6}$\pm$}0.0031\\
    pcl-v1 & 0.9434{\fontsize{6}{6}$\pm$}0.0020\\
    simclr-v1 & 0.9545{\fontsize{6}{6}$\pm$}0.0011\\
    byol & 0.9563{\fontsize{6}{6}$\pm$}0.0005\\
    swav & 0.9563{\fontsize{6}{6}$\pm$}0.0010\\
    barlow-twins & 0.9566{\fontsize{6}{6}$\pm$}0.0007\\
    sela-v2 & \textbf{0.9568{\fontsize{6}{6}$\pm$}0.0029}\\
    deepcluster-v2 & \textbf{0.9568{\fontsize{6}{6}$\pm$}0.0006}\\
    \shline
    \end{tabular}
    % }%
\end{center}
\end{table}

\newpage
\section{Backbone Architectures}
\noindent\textbf{ResNet18 and Resnet50~\cite{he2016deep}:}
One way to improve an architecture is to add more layers and make it deeper. Unfortunately, increasing the depth of a network do not work simply by stacking layers together. As a result, it can introduce the problem called vanishing gradient. Moreover, the performance might get saturated or decreased overtime. The main idea behind ResNet is to have identity shortcut connection which skips one or more layer. According to the authors, stacking layers should not decrease the performance of the network. The residual block allows the network to have identity mapping connections which prevents from vanishing gradient. The authors presented several versions of ResNet models including ResNet18, ResNet34, ResNet50 and ResNet101. The numbers indicate how many layers exist within the architecture. The more layers represent deeper network and the trainable parameters increase accordingly.

\noindent\textbf{ResNext50~\cite{xie2017aggregated}:}
In ResNext50, the authors introduced a new dimension C which is called Cardinality. The cardinality controls the size of the set of transformations addition to the dimensions of depth and width. The authors argue that increasing cardinality is more effective than going deeper or wider in terms of layers. They used this architecture in ILSVRC 2016 classification competition and secured the 2nd place. Comparing to ResNet50, ResNext50 has similar numbers of parameters for training and can boost the performance. In another word, ResNext50 could achieve almost equivalent performance to ResNet101 although ResNet101 has deeper layers. 

\noindent\textbf{DenseNet121~\cite{huang2017densely}:}
Increasing the depth of a network results in performance improvement. However, the problem arise when the network is too deep. As a result, the path between input and output becomes too long which introduces the popular issue called vanishing gradient. DenseNets simply redesign the connectivity pattern of the network so that the maximum information is flown. The main idea is to connect every layer directly with each other in a feed-forward fashion. For each layer, the feature-maps of all preceding layers are used as inputs, and its own feature-maps are used as inputs into all subsequent layers. According to the paper, the advantages of using DenseNet is that they alleviate the vanishing-gradient problem, strengthen feature propagation, encourage feature reuse, and substantially reduce the number of parameters. 

\noindent\textbf{Xception~\cite{chollet2017xception}:}
Xception network architecture was built on top of Inception-v3. It is also knows as extreme version of Inception module. With a modified depthwise separable convolution, it is even better than Inception-v3. The original depthwise separable convolution is to do depthwise convolution first and then a pointwise convolution. Here, Depthwise convolution is the channel-wise patial convolution and pointwise convolution is the 1x1 convolution to change the dimension. This strategy is modified for Xception architecture. In Xception, the depthwise separable convolution performs 1x1 pointwise convolution first and then channel-wise spatial convolution. Moreover, Xception and Inception-v3 has the same number of parameters. The Xception architecture slightly outperforms Inception-v3 on the ImageNet dataset and significantly outperforms Inception-v3 on a larger image classification dataset comprising 350 milions images and 17,000 classes.

\noindent\textbf{DRN-A-50~\cite{yu2017dilated}:}
Usually in image classification task the Convolutional Neural Network progressively reduces resolution until the image is represented by tiny feature-maps in which the spatial structure of the scene is not quite visible. This kind of spatial structure loss can hamper image classification accuracy as well as complicate the transfer of the model to a downstream task. This architecture introduces dilation which increases the resolutions of the feature-maps without reducing the receptive field of individual neurons. Dilated residual networks (DRNs) can outperform their non-dilated counterparts in image classification task. This strategy does not increase the model's depth or the complexity. As a result the number of parameters stays the same comparing to the counterparts. 

\noindent\textbf{SeNet154~\cite{hu2018squeeze}:}
The convolution operator enables networks to construct informative features by fusing both spatial and channel-wise information within local receptive fields at each layer. This work focused on a channel-wise relationship and proposed a novel architectural unit called Squeeze-and-Excitation (SE) block. This SE block adaptively re-calibrates channel-wise feature responses by explicitly modelling inter-dependencies between channels. These blocks can also be stacked together to form a network architecture (SeNet154) and generalise extremely effectively across different datasets. SeNet154 is one of the superior models used in ILSVRC 2017 Image Classification Challenge and won the first place. 

\noindent\textbf{SeResNet50, SeResNext50 and SeXception~\cite{hu2018squeeze}:}
The structure of the Squeeze-and-Excitation (SE) block is very simple and can be added to any state-of-the-art architectures by replacing components with their SE counterparts. SE blocks are
also computationally lightweight and impose only a slight increase in model complexity and computational burden. SE blocks were added to ResNet50 and ResNext50 model to design the new version. The pre-trained weights for SeResNet50 and SeResNext50 already exists where SeXception is not present. By adding SE blocks, we created the SeXception architecture and trained on ImageNet dataset to achieve the pre-trained weights. Later on we used the pre-trained weights for our transfer learning schemes.

%-------------------%

\section{Self Supervised Methods}

\noindent\textbf{InsDis~\cite{Wu2018insdis}:} 
\iffalse
InsDis treats each image as a distinct class and trains a non-parametric classifier to distinguish between individual classes based on noise-contrastive estimation (NCE)~\cite{gutmann2010noise}. InsDis introduces a feature memory bank maintaining a large number of noise samples (referred to as negative samples), to avoid exhaustive feature computing.
\fi
InsDis trains a non-parametric classifier to distinguish between individual instance classes based on NCE (noise-constrastive estimation)~\cite{gutmann2010noise}. Moreover, each instance of an image works as a distinct class of its own for the classifier. InsDis also introduces a feature memory bank to maintain a large number of noise samples (referring to negative samples). This helps to avoid exhaustive feature computing.
\noindent\textbf{MoCo-v1~\cite{He2020MocoV1} and MoCo-v2~\cite{chen2020improved}:} 
\iffalse
MoCo-v1 creates two views by applying two independent data augmentations to the same image $X$, referred to as positive samples. Like InsDis, the images other than $X$ are defined as negative samples stored in a memory bank. Additionally, a momentum encoder is proposed to ensure the consistency of negative samples as they evolve during training. Intuitively, MoCo-v1 aims to increase the similarity between positive samples while decreasing the similarity between negative samples. 
Through simple modifications inspired by SimCLR-v1~\cite{Chen2020Simple}, such as a non-linear projection head, extra augmentations, cosine decay schedule, and a longer training time to MoCo-v1, MoCo-v2 establishes a stronger baseline while eliminating large training batches.
\fi
MoCo-v1 uses data augmentation to create two views of a same image $X$ referring as positive samples. Similar to InsDis, images other than $X$ are defined as negative samples and they are stored in a memory bank. Moreover, to ensure the consistency of negative samples, a momentum encoder is introduced as the samples evolve during the training process. Basically, the proposed method aims to increase the similarity between positive samples while decreasing the similarity between negative samples. On the other hand, MoCo-v2 works similarly adding non-linear projection head, few more augmentations, cosine decay schedule, and a longer training time.

\noindent\textbf{SimCLR-v1~\cite{Chen2020Simple} and SimCLR-v2~\cite{chen2020big}:} 
\iffalse
SimCLR-v1 is proposed independently following the same intuition as MoCo. However, instead of using special network architectures (\eg a momentum encoder) or a memory bank, SimCLR-v1 is trained in an end-to-end fashion with large batch sizes.  Negative samples are generated within each batch during the training process. In SimCLR-v2, the framework is further optimized by increasing the capacity of the projection head and incorporating the memory mechanism from MoCo to provide more negative samples than SimCLR-v1.
\fi
The key idea of SimCLR-v1 is similar to MoCo yet proposing independently. Here, SimCLR-v1 is trained in an end-to-end fashion with larger batch sizes instead of using special network architectures (a momentum encoder) or a memory bank. Within each batch, the negative samples are generated on the fly. However, SimCLR-v2 optimizes the previous version by increasing the capacity of the projection head and incorporating the memory mechanism from MoCo to provide more meaningful negative samples.

\noindent\textbf{BYOL~\cite{grill2020bootstrap}:} 
\iffalse
Conventional contrastive learning methods such as MoCo and SimCLR relies on a large number of negative samples. As a result, they require either a large memory bank (memory consuming) or a large batch size (computational consuming).
% memory-consuming large memory bank or a computational-consuming large batch size. 
On the contrary, BYOL avoids the use of negative pairs by leveraging two encoders, named online and target, and adding a predictor after the projector in the online encoder. BYOL thus maximizes the agreement between the prediction from the online encoder and the features computed from the target encoder. The target encoder is updated with the momentum mechanism to prevent the collapsing problem.
\fi
MoCo and SimCLR methods mainly relies on a large number of negative samples and they require either a large memory bank or a large batch size. On the other hand, BYOL replaces the use of negative pairs by adding online encoder, target encoder and a predictor after the projector in the online encoder. Both the target encoder and online encoder computes features. The key idea is to maximize the agreement between target encoder's features and prediction from online encoder. To prevent the collapsing problem, the target encoder is updated by the momentum mechanism.

\noindent\textbf{PIRL~\cite{Misra2020Self}:} 
\iffalse
Instead of using instance discrimination like InsDis and MoCo, PIRL adapts Jigsaw and Rotation as proxy tasks. Specifically, the positive samples are generated by applying Jigsaw shuffling or rotating by \{$0^{\circ}$, $90^{\circ}$, $180^{\circ}$, $270^{\circ}$\}. PIRL defines loss function based on noise-contrastive estimation (NCE) and uses a memory bank following InsDis. In this paper, We only benchmarks PIRL with Jigsaw shuffling, which yields better performance than its rotation counterparts.
\fi
Both InsDis and MoCo takes the advantage of using instance discrimination. However, PIRL adapts the Jigsaw and Rotation as proxy tasks. Here, the positive samples are generated by applying Jigsaw shuffling or rotating by \{$0^{\circ}$, $90^{\circ}$, $180^{\circ}$, $270^{\circ}$\}. Following InsDis, PIRL uses Noise-Constrastive estimation (NCE) as loss function and  a memory bank. 

\noindent\textbf{DeepCluster-v2~\cite{caron2021swav_appendix}:}
\iffalse
DeepCluster~\cite{caron2018deep} learns features in two phases: (1) self-labeling, where pseudo labels are generated by clustering data points using the prior representation--- yielding cluster indexes for each sample; (2) feature-learning, where the cluster index of each sample is used as a classification target to train a model. The two phases are performed repeatedly until the model converges. Rather than classifying the cluster index, DeepCluster-v2 explicitly minimizes the distance between each sample and the corresponding cluster centroid. DeepCluster-v2 finally applies stronger data augmentation, a MLP projection head, a cosine decay schedule, and multi-cropping to improve the representation learning. 
\fi
DeepCluster~\cite{caron2018deep} uses two phases to learn features. First, it uses self-labeling, where pseudo labels are generated by clustering data points using prior representation yielding cluster indexes for each sample. Secondly, it uses feature-learning, where each sample's cluster index is used as a classification target to train a model. Until the model is converged, the two phases mentioned above is performed repeatedly. The DeepCluster-v2 minimizes the distance between each sample and the corresponding cluster centroid. DeepCluster-v2 also uses stronger data augmentation, MLP projection head, cosine decay schedule, and multi-cropping to improve the representation learning.

\noindent\textbf{SeLa-v2~\cite{asano2020selflabelling}:}
\iffalse
Similar to clustering methods, SeLa~\cite{asano2020selflabelling} requires a two-phase training (\ie self-labeling and feature-learning). However, instead of clustering the image instances, SeLa formulates self-labeling as an optimal transport problem, which can be effectively solved by adopting the Sinkhorn-Knopp algorithm. Similar to DeepCluster-v2, the updated SeLa-v2 applies stronger data augmentation, a MLP projection head, a cosine decay schedule, and multi-cropping to improve the representation learning.
\fi
SeLa also requires two-phase training (\ie self-labeling and feature-learning). SeLa focuses on self-labeling as an optimal transport problem and solves it using Sinkhorn-Knopp algorithm. SeLa-v2 also uses stronger data augmentation, MLP projection head, cosine decay schedule, and multi-cropping to improve the representation learning.

\noindent\textbf{PCL-v1 and PCL-v2~\cite{asano2020selflabelling}:} 
PCL-v1 aims to bridge contrastive learning with clustering. PCL-v1 adopts the same architecture as MoCo, including an online encoder and a momentum encoder. Following clustering-based feature learning, PLC-v1 also uses two phases (self-labeling and feature-learning). The features obtained from the momentum encoder are clustered in self-labeling phase. On the other hand, PCL-v1 generalizes the NCE loss to ProtoNCE loss instead of classifying the cluster index with regular cross-entropy. This was done in PCL-v2 as an improvement step.

\iffalse
\noindent\textbf{SwAV~\cite{caron2021swav_appendix}:}
SwAV takes advantage of both contrastive learning and clustering. Like SeLa, SwAV calculates prototype assignment (or codes) for each data sample with the Sinkhorn-Knopp algorithm. However, SwAV performs assignments at the batch level instead of epoch level and thus works online. Compared with contrastive learning such as MoCo and SimCLR, SwAV ``swapped" predicts the codes obtained from one view using the other view rather than comparing their features directly. Additionally, SwAV proposes a multi-crop strategy, which can be adopted by other methods and consistently improve their performance.

SwAV takes advantages of both contrastive learning and clustering techniques. Similar to SeLa, SwAV calculates cluster assignments (codes) for each data sample with the Sinkhorn-Knopp algorithm. However, SwAV performs online cluster assignments, \ie at the batch level instead of epoch level. Compared with contrastive learning approaches such as MoCo and SimCLR, SwAV ``swapped" predicts the codes obtained from one view using the other view rather than comparing their features directly. Additionally, SwAV proposes a multi-cropping strategy, which can be adopted by other methods to consistently improve their performance.
\fi

\noindent\textbf{SwAV~\cite{caron2021swav_appendix}:} SwAV uses both constrastive learning as well as clustering techniques. For each data sample, SwAV calculates cluster assignments (codes) with the help of Sinkhorn-Knopp algorithm. Moreover, SwAV works online performing assignments at the batch level instead of epoch level.
\iffalse
\noindent\textbf{InfoMin~\cite{tian2020makes}:} 
InfoMin hypothesizes that good views (or positive samples) should only share label information w.r.t the downstream task while throwing away irrelevant factors, which means optimal views for contrastive representation learning are task-dependent. Following this hypothesis, InfoMin optimizes data augmentations by further reducing mutual information between views.
\fi

\noindent\textbf{InfoMin~\cite{tian2020makes}:} 
InfoMin suggested that for contrastive learning, the optimal views depend upon the downstream task. For optimal selection, the mutual information between the views should be minimized while preserving the task-specific information. 

\iffalse
\noindent\textbf{Barlow Twins~\cite{zbontar2021barlow_appendix}:}
function by measuring the cross-correlation matrix between the outputs of two online encoders fed with two random augmented views of the same sample. The representations are learned by making the cross-correlation matrix as close to the identity matrix as possible. As a result, the similarity between representations of two views is maximized, sharing the same goal as contrastive learning. On the other hand, the redundancy between the components of two representations is minimized. 
\fi
\noindent\textbf{Barlow Twins~\cite{zbontar2021barlow_appendix}:}  The Barlow Twins consists of two identical networks fed with the two distorted versions of the input sample.
The network is trained such that the cross-correlation matrix between the two resultant embedding vectors is close to the identity. A regularization term is also included in the objective function to minimize redundancy between embedding vectors' components.

\end{document}